# The ASIM Mission on the International Space Station

Torsten Neubert[1] · Nikolai Østgaard[2] · Victor Reglero[3] · Elisabeth Blanc[4] ·
Olivier Chanrion[1] · Carol Anne Oxborrow[1] · Astrid Orr[5] · Matteo Tacconi[5] ·
Ole Hartnack[6] · Dan D.V. Bhanderi[6]

**Abstract** The Atmosphere-Space Interactions Monitor (ASIM) is an instrument suite on the International Space Station (ISS) for measurements of lightning, Transient Luminous Events (TLEs) and Terrestrial Gamma-ray Flashes (TGFs). Developed in the framework of the European Space Agency (ESA), it was launched April 2, 2018 on the SpaceX CRS-14 flight to the ISS. ASIM was mounted on an external platform of ESA's Columbus module eleven days later and is planned to take measurements during minimum 3 years. The instru-

The ASIM mission on the International Space Station

✉ T. Neubert
neubert@space.dtu.dk

N. Østgaard
Nikolai.Ostgaard@ift.uib.no

V. Reglero
victor.reglero@uv.es

E. Blanc
Elisabeth.Blanc@cea.fr

O. Chanrion
chanrion@space.dtu.dk

C.A. Oxborrow
cao@space.dtu.dk

A. Orr
Astrid.Orr@esa.int

M. Tacconi
Matteo.Tacconi@esa.int

O. Hartnack
oh@terma.com

D.D.V. Bhanderi
dbh@terma.com

[1] National Space Institute, Technical University of Denmark (DTU Space), Kongens Lyngby, Denmark

[2] Birkeland Center for Space Science, University of Bergen, Bergen, Norway

ments are an x- and gamma-ray monitor measuring photons from 15 keV to 20 MeV, and an array of three photometers and two cameras measuring in bands at: 180–250 nm, 337 nm and 777.4 nm. Additional objectives that can be addressed with the instruments relate to space physics like aurorae and meteors, and to Earth observation such as dust- and aerosol effects on cloud electrification. The paper describes the scientific objectives of the ASIM mission, the instruments, the mission architecture and the international collaboration supported by the ASIM Science Data Centre. ASIM is the first space mission with a comprehensive suite of instruments designed to measure TLEs and TGFs. Two companion papers describe the instruments in more detail (Østgaard et al. in Space Sci. Rev., 2019; Chanrion et al. in Space Sci. Rev., 2019).

## 1 A Brief History of ASIM

Launched on April 2, 2018, ASIM carries dedicated instruments designed to measure Transient Luminous Events (TLEs) and Terrestrial Gamma-ray Flashes (TGFs) generated by the electrical activity of thunderstorms. TLE is the common name for glimpses of light in the stratosphere and mesosphere above thunderstorms. They include electrical discharges such as sprites, jets and gigantic jets, and luminous excitation of the atmosphere such as the elves. TGFs are bursts of bremsstrahlung from energetic particle beams accelerated in thunderstorm processes. Most of the manifestations of TLEs and TGFs were discovered in the period 1989–2002 (Franz et al. 1990; Wescott et al. 1995; Fishman et al. 1994; Fukunishi et al. 1996; Pasko et al. 2002). A schematic overview of the main entities is shown in Fig. 1.

The history of ASIM dates back to the time of the discoveries of the new thunderstorm manifestations. In 1997, the space science community in Denmark hoped to establish a national small-satellite program based on the large interest in Denmark's first satellite Ørsted, launched in 1999 for high-precision measurements of Earth's magnetic field (a forerunner of the ESA *Swarm* mission). In response to a call for ideas, issued by the Small Satellite Program Committee established by the Ministry for Research, a Danish consortium proposed the Atmospheric X-ray Observatory (AXO), which was selected for the short-list of mission proposals. AXO was to use Danish expertize in x- and gamma-ray instrumentation to orbit a dedicated instrument that would be aimed at the Earth's atmosphere for detection of TGFs. However, the funding for a national programme did not materialize and Denmark has not launched a nationally funded satellite since Ørsted.

In 2003, the Human Spaceflight Micro-gravity and Exploration Directorate (HME) of ESA invited ideas for external payloads on the Columbus module of the ISS. DTU Space felt that the ISS was ideal for an AXO-type investigation because its orbit is the lowest available and covers most of Earth's thunderstorm regions. It was decided to propose AXO, which was later resubmitted formally to an announcement of opportunity by the HME directorate

[3]  Image Processing Laboratory, University of Valencia, Valencia, Spain

[4]  CEA DAM DIF, 91297 Arpajon, France

[5]  Directorate of Human Spaceflight and Robotics Exploration, European Space Research and Technology Center, Noordwijk, The Netherlands

[6]  Terma, Space Division, Herlev, Denmark

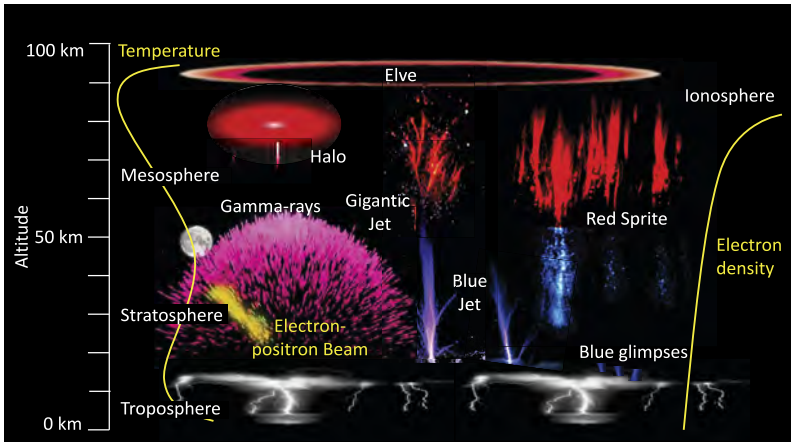

**Fig. 1** The zoo of upper atmospheric phenomena powered by thunderstorms. Terrestrial Gamma-ray Flashes (TGFs) are flashes of gamma-rays from thunderstorms generated by bremsstrahlung from bursts of energetic electrons. These photons reach energies that allow for pair production of positrons and electrons. Transient Luminous Emissions (TLEs) are electrical discharges that include blue glimpses at the top of thunderstorms, blue jets, gigantic jets and red sprites. TLEs also include elves, the rapidly expanding rings of emissions at the bottom ionosphere, and the halos. Credit DTU Space; TGF: NASA

under the name of ASIM. It received the evaluation "Excellent" and was included in ESA's portfolio Research Programme for the ISS.

An Instrument Pre-Phase A study was completed in June 2004 with the recommendation from ESA to enhance the instrumentation and to enlarge the consortium to a broader European collaboration. Two groups were then invited to join, both with solid track records in x-ray instrumentation. One group is at the University of Bergen that participated in the PIXIE instrument (Polar Ionospheric X-ray Imaging Experiment) on the POLAR spacecraft and the IBID instrument (Imager on Board the Integral Satellite) on INTEGRAL (International Gamma-Ray Astrophysics Laboratory), and with many years of experience with x-ray instruments on balloons and rockets. The other group is at the University of Valencia, that participated in the JEM-X (Joint European X-Ray Monitor) on INTEGRAL. Additional partners that entered the consortium was OHB-Italy with a payload computer for interface to the ISS (flown on an earlier mission) and Space Research Center, Polish Academy of Science, with a power supply for the x- and gamma-ray instrument. The technical consortium was led by Terma Space, Denmark. A study of an enhanced design using ESA's Concurrent Design Facility (CDF) was performed in October 2004 at ESA's European Space Technology Centre (ESTEC), which was followed by a Payload Phase A study 2005–2006. Phase B was conducted during 2007–2009 and Phase C/D during 2010–2018.

In parallel with ASIM, French scientists proposed the TARANIS satellite (Tool for the Analysis of RAdiations from lightNIngs and Sprites) to the French National Space Agency (CNES), also for studies of thunderstorms, TLEs and TGFs, which was selected in 2001 for a pre-phase A study. To validate the new concept of optical nadir observations used in TARANIS, the French team flew the LSO (Lightning and Sprite Observations) experiment composed of dual-band test cameras on the ISS missions Andromede (2001), Odissea (2002) and Delta (2004), and demonstrated for the first time that spectral observations exploiting atmospheric transmission properties can separate the signatures of sprites from the simultaneous lightning illumination of the cloud below (Blanc et al. 2004). This important result also guided the design of the ASIM observational strategy (see below).

Simultaneously with our efforts to realize ASIM and TARANIS, French and Danish scientists began building European expertize in this, at the time, rather new field of thunderstorm science. Thus, the first observations of sprites over Europe were taken from Observatoire Midi-Pyrenees, documenting that the weaker European thunderstorms also generate sprites (Neubert et al. 2001). Funding was secured for a European-wide research training network CAL in the European Commission FP5 programme (Coupling of Atmospheric Layers, 2002–2006) of 11 groups and 8 young scientists to study lightning and TLEs (Neubert et al. 2008). During the four years of the network, the groups intensified observations from Southern France, forming the loosely organized *EuroSprite* observational network (e.g. Soula et al. 2014; Arnone and Dinelli 2016) and organizing two very successful summer schools in Corte, Corsica (Fullekrug et al. 2004). The success of the CAL network gave momentum to the ASIM mission by demonstrating the commitment and expertize of European research groups backing the mission.

TARANIS, named after the Gaelic thunder god, was selected to phase A in 2004 after the successful ISS demonstration experiments, and will be launched. The teams behind ASIM and TARANIS have succeeded in securing two space missions with complementary instruments because of the close collaboration of European scientists. The collaboration intensified over the years via the European Science Foundation (ESF) network TEA-IS (Thunderstorm Effects on the Atmosphere-Ionosphere System, 2010–2016) and the European Commission network SAINT (Science and INnovation with Thunderstorms, 2017–2021) of 23 partners and 15 PhD students to conduct research in ASIM/TARANIS science.

## 2 Research Objectives

### 2.1 Introduction

Processes in the atmosphere can affect human lives, like severe storms that carry torrential rain, devastating hail or deadly lightning, or on longer time scales by changing the climate. The atmosphere is also a screen that makes phenomena in outer space visible, for instance as aurorae that are caused by particles accelerated by the energy transferred from the solar wind interacting with the Earth's magnetosphere. For many years, this dual role was considered to be played by two distinct characters, with weather and climate taking place in the troposphere below $\sim 15$ km altitude and space processes displayed in the mesosphere above $\sim 50$ km. Consequently, the study of these aspects was conducted by two separate scientific communities. However, these boundaries have shifted during the past several years. For instance, to understand weather and climate, we must understand the exchange of greenhouse gasses between the troposphere and the stratosphere mediated by atmospheric inertial gravity waves and thunderstorm convection (e.g. Solomon et al. 2010; Fritts and Alexander 2003). The roles themselves are also changing. Space processes are not just displayed by the atmosphere, they are also thought to affect changes in the weather and climate, for instance through perturbations to greenhouse gas agents by energetic particle precipitation carried to low altitudes by the Dobson circulation (e.g. Funke et al. 2005; Randall et al. 2005) and modification to cloud cover by cosmic rays (e.g. Svensmark et al. 2016). With the discovery in 1990 of an electrical discharge in the mesosphere powered by thunderstorms below, the sprite, the paradigm of separate dual roles broke down. We can no longer assume the mesosphere is reserved for passive displays of space processes, in fact, we now must consider the stratosphere and mesosphere as a playground for interacting atmospheric and space processes, thus the name for the ASIM mission.

The mesosphere and lower thermosphere are the regions of the atmosphere that are least known. They are too low for in-situ spacecraft observations, too high for balloon observations, and remote sensing is hampered by low densities and a high degree of variability over a range of temporal and spatial scales. As a consequence, these regions are understood primarily on large spatial and temporal scales typical of planetary waves and internal gravity waves (spread over several km and several minutes). However, the complex spatial structure of sprites suggest that the mesosphere is inhomogeneous at smaller scales, perhaps induced by meteors and cosmic rays. Understanding the physics of TLEs, therefore, holds the promise of using TLE observations as probes of the small-scale structure of the stratosphere, mesosphere and lower thermosphere.

The core objectives which drove the ASIM instrument design relate to TLEs, TGFs and lightning. They are:

– To conduct a comprehensive global survey of TLEs and TGFs covering all local night times and seasons
– To secure data for understanding the fundamental kinetic processes of TLEs and TGFs
– To understand the relationship of TLEs and TGFs to lightning activity

The ASIM observations will improve estimations of the regional effects on the atmosphere of these phenomena and will give us new insight into the fundamental processes underpinning lightning. The temporal and spatial scales of discharges are approximately proportional to the neutral density as $1/n_n$ and discharges are, therefore, many orders of magnitudes slower and larger in the stratosphere and mesosphere relative to ground levels, which brings them within reach of today's instrumentation. In a sense, measurements of TLEs allow us to look inside lightning. Likewise, TGFs are signatures of processes inside lightning that we have not been able to measure before. The new understanding of lightning that can be gained by studying TLEs and TGFs allows us to get at the kinetic physics of the lightning process, a step that is needed to accurately model the perturbations to ozone and NOx concentrations and thus to quantify global effects of lightning discharges.

Additional objectives that can be addressed with the ASIM observations are processes related to space science:

– Meteor precipitation in the atmosphere
– The optical and x-ray aurorae

and topics related to Earth observation:

– The effects on cloud electrification of dust storms, forest fires and volcanoes
– Intensification of hurricanes and its relation to eye-wall lightning activity

The mission's objectives cover a wide range of scientific fields which require a broad, cross-disciplinary collaboration of scientists. Furthermore, simultaneous observations from other instruments in space and on the ground are important, as well as numerical models, for instance of TLEs, TGFs and of their atmospheric perturbations. For these reasons, the ASIM mission is open for international collaboration.

### 2.2 Some Background on TLEs and TGFs

Transient Luminous Emissions (TLEs) are glimpses of optical radiation in the stratosphere, mesosphere and lower thermosphere above thunderstorms. The first scientific report came in 1990 (Franz et al. 1990) which presented an image of a large luminous flash in the mesosphere above a distant thunderstorm. The image was in black and white and the luminous region appeared in a single frame of a video sequence. The discovery led to quite some activity in the scientific community, both to understand why such events had not been discovered earlier and to take more observations. Indeed, it was found that the space shuttle orbiter cameras had recorded similar events (Boeck et al. 1998), which was realized around the same time. In addition, it turned out that eyewitness accounts existed dating back to the 19th century (Lyons et al. 2003) and that discharges to the ionosphere was proposed in the 1920'ies by C.T.R. Wilson (Williams 2010). A major step forward followed with a NASA aircraft campaign where the first images in color were taken. They showed that the dominant color was red (Sentman et al. 1995), and discovered in addition blue jets shooting up from the top of thunderstorm clouds and into the stratosphere (Wescott et al. 1995). Such events, of lightning reaching into the stratosphere, were previously reported in Vaughan and Vonnegut (1989) and references therein, however the images of the NASA campaign were the most spectacular at that time. Following these observations, the names of 'red sprites' and 'blue jets' were coined and the common name of TLEs was adopted for all thunderstorm-related emissions in the upper atmosphere.

During the same time period, the first report of gamma-ray flashes from thunderstorms appeared. They were observed by the BATSE instrument on the CGRO satellite, which was designed to measure Cosmic Gamma-ray Bursts, but had also detected shorter-duration bursts from the Earth as the satellite passed over thunderstorms. The phenomena was named accordingly as Terrestrial Gamma-ray Flashes (Fishman et al. 1994). TGFs are bremsstrahlung from energetic electrons accelerated in thunderstorm's electrical fields. Although the discovery of TGFs was a surprise, high-energy runaway electrons had been suggested by Wilson in 1929 (Williams 2010).

Since the early days, a variety of TLEs have been documented. They include elves (Fukunishi et al. 1996), gigantic jets (Pasko et al. 2002), trolls, gnomes and pixies (Lyons et al. 2003), and blue glimpses (Chanrion et al. 2017). Global observations of TLEs have most notably been taken by the ISUAL instrument (Imager of Sprites and Upper Atmospheric Lightning) on the FormoSat-2 polar orbiting satellite, viewing the Earth's atmosphere towards the limb (Frey et al. 2016). The observations allowed for studies of the statistics and distribution of TLEs (Chen et al. 2008), and for single event studies of their dynamics using a complement of array photometers (e.g. Kuo et al. 2009). Additional observations have been conducted from the space station by the Japanese GLIMS (Global Lightning and sprIte MeasurementS) experiment in the nadir-viewing geometry (Sato et al. 2015). We now know that sprites are $\sim 10$ ms-duration discharges in the mesosphere caused by the quasi-electrostatic electric field, usually following a positive cloud-to-ground lightning flash in the storm below (Pasko et al. 1997, 1998). Blue jets and gigantic jets are upward lightning typically lasting 100 ms to 1 sec caused by charge imbalances in the top layer of clouds (Williams 2008; Krehbiel et al. 2008). Elves are $\sim 1$ ms-duration excitation of neutral atmospheric species by free electrons in the bottom ionosphere energized by the cloud electromagnetic pulse (Barrington-Leigh and Inan 1999). Several review papers have been published on the observation and physics of TLEs, e.g. Neubert (2003), Neubert et al. (2008), Pasko et al. (2012), Lyons (2006), Pasko (2010).

For TGFs, RHESSI (Reuven Ramaty High Energy Solar Spectroscopic Imager) (Smith et al. 2005), AGILE (Italian Astro-rivelatore Gamma a Immagini Leggero satellite) (Marisaldi et al. 2014) and Fermi (Fermi Gamma-ray Space Telescope) (Briggs et al. 2013) have provided energy spectra with improved energy range and resolution, have documented the existence of positron beams created through photon pair production (Briggs et al. 2011) and have allowed for detailed analysis of the relationship between TGFs and lightning processes (Østgaard et al. 2013; Cummer et al. 2014). The global coverage of these missions extends to ±30° latitude.

There are two classes of electron acceleration mechanisms that are currently debated. One considers a large region of the cloud where the electric field drives relativistic run-away electron avalanches seeded by electrons with energies in the run-away regime, for instance created by cosmic ray ionization of the atmosphere (Gurevich et al. 1992) with positron feedback (Dwyer 2012). In the other theory, electrons are accelerated in a smaller region of high electric field around the lightning leader tip (Carlson et al. 2010; Celestin and Pasko 2011) and in its streamer corona (Chanrion and Neubert 2010; Babich et al. 2017). It is likely that both mechanisms operate in thunderstorms. A comprehensive review of observations and theories is given in Dwyer et al. (2012).

### 2.3 The New Observations with ASIM

ASIM is the first space mission that carries x- and gamma-ray instrumentation designed to observe TGFs. The detectors are optimized for the dominant part of the energy range of TGFs and have a sufficient area that allows characterization of the energy distribution of individual TGFs rather than average distributions of many TGFs as done in past missions. In addition, the detector trigger logic that identifies events is designed to capture the ms-duration events which, in combination with the larger detector area, makes it the most sensitive TGF detector flown in space. Whereas the source of TGFs in the past has been determined indirectly from time correlations of lightning within the region below the satellites, the ASIM detector has the capability to determine the direction of arrival of the flash photons, allowing us to pinpoint their source location. The TGF observations are supported, for the first time, by a comprehensive suite of optical instruments that detect lightning and TLEs at higher temporal and spatial resolution than past experiments such as ISUAL and GLIMS. We expect that the simultaneous optical observations will allow us to determine the characteristics of the lightning processes in which TGFs are generated.

The superior resolution of the optical measurements will help us to understand the electrical properties of thunderstorms that make them active sources of the various forms of TLEs, in particular in the blue band and at the top of thunderstorm clouds and into the stratosphere. For instance, the processes of blue glimpses on the top of thundercloud turrets, blue jets reaching into the stratosphere and gigantic jets to the ionosphere are still relatively poorly understood. We suggest that more data, as expected from ASIM, are required to consolidate the models of the charge distributions in clouds that generate these events and to understand their polarity. We also expect that the ASIM observations will allow us to study the leader-streamer properties gigantic jets.

## 3 Mission Architecture

### 3.1 Measurement Concept

The approach has been to develop an x- and gamma-ray detector, the MXGS (Modular X- and Gamma-ray Sensor) that is sensitive to TGFs, and to support these measurements by a suite of optical sensors that measure simultaneous lightning activity in the clouds and TLE activity above them. The optical sensors form the MMIA (Modular Multi-spectral Imaging

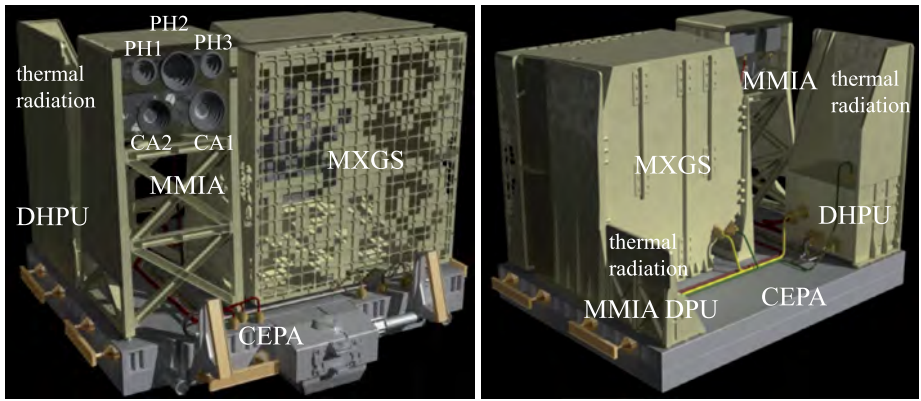

**Fig. 2** The ASIM payload

Array). Other instruments were considered such as electromagnetic wave sensors, however, the ISS was thought to be a too noisy environment for such measurements.

The ISS is orbiting in the local vertical/local horizontal (LVLH) attitude frame (as an airplane). The MXGS on the ISS is pointing towards the nadir to maximize the number of photons detected from a TGF. In the nadir-viewing geometry, atmospheric absorption is minimized and the source regions are as close to the detector as possible. The MMIA sensors view the same region of the atmosphere in order to allow correlation of the observations. While the orientation is ideal for support of the TGF observations of the MXGS, the challenge for the MMIA sensors is to distinguish TLEs from lightning, and even more difficult, to identify TLEs when they are viewed on top of lightning-illuminated clouds. Here, we use the principle demonstrated in the LSO experiment on the ISS (Blanc et al. 2004), that is, to measure in a band that is little affected by the atmosphere and bands that are absorbed in the atmosphere to varying degrees. High-altitude emissions are less damped than low-altitude ones in these bands, and therefore the relative amplitude in the different bands contains information on the altitude of the emissions.

### 3.2 The Payload

The payload with the instruments is shown in Fig. 2. The MXGS and the MMIA are mounted on the CEPA (Columbus External Payload Adapter) together with the ASIM payload computer, the DHPU (Data Handling and Power Unit), and the MMIA instrument computer (Data Processing Unit—DPU). The DHPU is behind the MMIA and hidden by an extended radiator plate, and the MMIA DPU is behind the MXGS. The MXGS DPU is a derivative of the MMIA DPU and is mounted inside the MXGS together with the MXGS power supply. The MMIA power supply is integrated with the MMIA DPU. The CEPA provides the structural support for the payload, and power and data connections from/to the DHPU and the ISS. The DHPU converts the 120 V power supply from the ISS to 28 V instrument supply and handles all data communication. The payload mass is 314 kg and its dimensions are $122 \times 134 \times 99$ cm$^3$. The power consumption is 200 W, on top of which 230 W are allocated for thermal heaters. ASIM has a down-link allocation of 200 kbps continuous data, which is fully utilized since the instruments collect vast amounts of data, and low prioritized data are not always able to be down-linked.

Table 1 MXGS instrument specifications

| MXGS | LED | HED |
|---|---|---|
| Geometrical area (cm$^2$) | 1024 | 900 |
| Energy range | 15–400 keV | 200 keV–20 MeV |
| Energy resolution | < 10% @ 60 keV | < 15% @ 662 keV |
| Angular resolution point source | < 0.7° | |
| Relative time accuracy | 10 μs | 10 μs |
| Sensitivity (signal/noise) | > 7 | > 15 |

The main industrial contractor was Terma Space, Denmark. Terma was also the main subcontractor of the MMIA and delivered the optical cameras. DTU Space developed the photometers, the structure and the MMIA DPU with integrated power supply. The main subcontractor for the MXGS was DTU Space who also developed the MXGS DPU. The MXGS detectors were supplied by University of Bergen and the structure and coded mask by University of Valencia. The MXGS power supply unit was supplied by the Space Research Center, Polish Academy of Science and the DHPU by OHB-Italy.

### 3.3 The Modular X- and Gamma-ray Sensor (MXGS)

The MXGS has two detector layers, a low energy detector (LED) and behind it, a high-energy detector (HED). The LED is made of CdZnTe crystals in the detector plane giving $128 \times 128$ pixels. At the aperture is a coded mask that absorbs low-energy photons. This produces a shadow pattern in the detector plane when photons arrive from a localized source, and from that pattern, the direction of arrival can be determined. The HED is a layer of BGO scintillators coupled to photomultiplier tubes. The detector photons arrive primarily through the aperture and pass through the coded mask and the LED detector plane. Photons arriving from the sides and back are stopped by shielding. The combination of the two detectors covers photon energies from 15 keV to 20 MeV. With LED and HED geometric detector areas of 1024 cm$^2$ and 900 cm$^2$, the sensitivity is expected to allow for estimation of individual TGF spectra and source. Some instrument specifications are given in Table 1 and a more detailed description can be found in the companion paper (Østgaard et al. 2019).

### 3.4 The Modular Multispectral Imaging Array (MMIA)

The spatial resolution of optical events is provided by a pair of cameras, one in the near-UV at the strongest spectral line of the nitrogen 2nd positive system (337 nm) (N$_2$ : C$_3\Gamma_u \rightarrow$ B$_3\Gamma_g$)(0, 0) and the other in the strongest lightning emission band, OI (777.4 nm) (Christian et al. 1989, 2003). The temporal resolution is given by three photometers sampling at 100 kHz in the UV at 180–230 nm, capturing part of the nitrogen Lyman-Birge-Hopfield (LBH) band, and in the near-UV at 337 nm and lightning band at 777.4 nm. While some lightning emissions in the UV and near-UV will reach the sensors, the signals will be damped relative to those in the lightning band. For example, the transmission from the thunderstorm cloud tops at 15 km altitude is below $\sim 0.001\%$ in 180–230 nm. At 337 nm it is $\sim 75\%$ from 15 km altitude and $\sim 50\%$ from 10 km altitude (Neubert and Chanrion 2010).

The UV photometer will give no signal from lightning itself, except possibly for very strong lightning in the upper levels of clouds, and it is used as the primary sensor that tells if an event includes a TLE. We note here, that the waveband sensitivity of the UV photometer does not rely on a filter but rather on the cathode response (upper band limit) and the quartz glass window (lower limit). There is therefore no possibility that signals at other wavelengths will contaminate the measurements. If an event is likely to be a TLE, the spatial and temporal structure of emissions will aid in the interpretation of the kind of TLE was observed. The higher temporal and spatial resolution of the ASIM measurements relative to past observations (Blanc et al. 2004; Sato et al. 2015; Frey et al. 2016) are expected to make this task simpler and the conclusions more robust. Some specifications of the MMIA sensors are given in Table 2 and in Chanrion et al. (2019).

Table 2 MMIA instrument specifications

| MMIA | Cameras | Photometers |
|---|---|---|
| FOV (nadir) diagonal/diameter | 80° | 80° |
| Pixels | 1024 × 1024 | |
| Spatial resolution (ground) | 400–500 m | |
| Temporal resolution | 83 ms | 10 μs |
| Relative time accuracy | 10 μs | 10 μs |
| Spectral bands (nm) (center/width) | CA1: 337/5 | PH1: 337/5 |
| | | PH2: 180–230 |
| | CA2: 777.4/3 | PH3: 777.4/5 |
| Sensitivity (ph/m$^2$/s) Flux at aperture (CA1, 2 single pixel) | CA1: $3.2 \times 10^6$ | PH1: $1.5 \times 10^{12}$ |
| | | PH2: $6.9 \times 10^{12}$ |
| | CA2: $4.2 \times 10^7$ | PH3: $2.2 \times 10^{12}$ |

### 3.5 Orbit

The ASIM instruments are mounted on an external platform of the Columbus module. The location is on the starboard/ram side relative to the ISS flight direction. The inclination of the ISS orbit is 51.6°, which brings the instruments over all the major thunderstorm regions of the Earth. The orbit plane drifts 6° westward per day, thereby covering all local times (LT) in about 2 months. The minimum duration of the mission is two years, which allows complete coverage of all seasons. The ISS is the lowest available platform in space with altitudes from 370 to 460 km, which brings the instruments as close as possible to the thunderstorm activity to be measured. The ISS is therefore ideal for Earth observation objectives such as global characterization of thunderstorms. The location of ASIM on the Columbus module is shown in Fig. 3.

### 3.6 Instrument Modes

The instrument sensors of the MXGS and MMIA are controlled by two computers based on the Xilinx Virtex5 FPGA, being flown for the first time by ESA. The primary mode of the sensors is the trigger mode, where the instruments run continuously and sudden increases in any sensor signal level is detected by the instrument application software which saves the trigger data from all sensors to memory. In the case of MMIA, a minimum of three camera frames are selected, the frame in which a trigger was detected and one frame before and after, and photometer data corresponding to the exposure of the frames. This strategy of on-board data selection greatly reduces the amount of down-link data. Further reduction is achieved by on-board cropping of the image data to select only the part of the frame with optical signal, followed by lossless image compression. The MMIA sensors are so light sensitive that they run only during the night. The MMIA computer crops and compresses the selected data during the day. The MXGS LED and HED also have a trigger mode. The data stored by them is 1 s second before and 1 s after the trigger. The data stored and down-linked is the deposition energy, time of arrival and the detecting sensor element of each photon. The MXGS HED detector runs day and night and the LED only at night because of daylight contamination. The HED is off in the South Atlantic Anomaly and the LED enters auroral mode (see below).

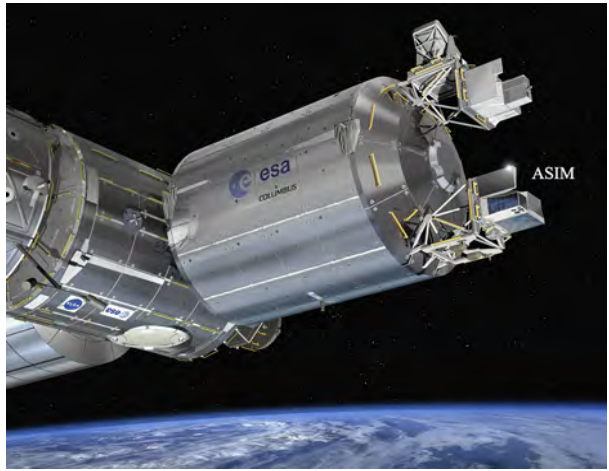

**Fig. 3** ASIM on the Columbus laboratory module viewing downwards towards the atmosphere

The MXGS and MMIA can trigger each other, thereby securing a complete data set of all sensors during a trigger event. The trigger data are assigned priorities according to the combination of sensors that simultaneously trigger. For example, if only one MMIA sensor triggers, it is likely to be caused by an energetic particle and the trigger will be ignored. If all the optical sensors trigger except the UV photometer, the event is likely a lightning event. In this case, the MMIA photometer data and the row and column sums of the MMIA camera frames are assigned priority 1 and the camera raw pixel data priority 3. If more than one MMIA sensor trigger simultaneously with the LED or HED, the event is possibly a TGF and the data will be given priority 1. MMIA by itself generates priority 1 data if two or more MMIA sensors trigger and one of these is the UV photometer, then the event is possibly a TLE. Data are down-linked with the highest priority first. The low priority data may be overwritten if sufficient memory is not available.

In addition to the trigger mode, the MMIA can be commanded in timed observation mode where continuous recordings are taken during a limited period in time, typically with longer exposure times or lower frame rates than for triggered observations. The timed mode is used for observations of aurora or special localized targets such as forest fires, volcanic eruptions etc. The MXGS has an auroral mode where photons are binned in energy and time. If the photon count rate exceeds a threshold, the instrument will automatically change from trigger mode to auroral mode. More information on the MMIA and MXGS instruments is given in Østgaard et al. (2019), Chanrion et al. (2019).

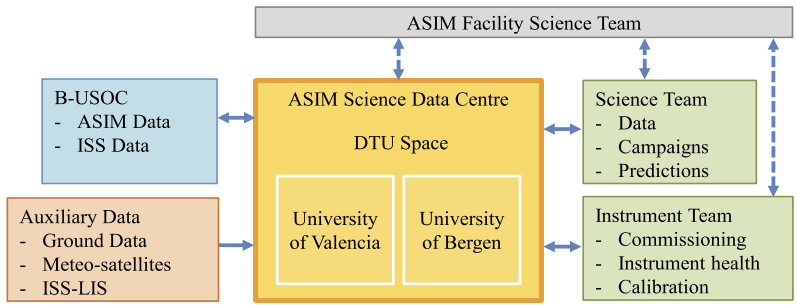

**Fig. 4** A schematic representation of the ASDC. On the left is the data exchange with external data providers and on the right, the data exchange with the instrument teams and the science users

## 4 ASIM Organisation and Data Access

The ASIM Facility Science Team (FST) was established by ESA to advise ESA on the development and operation of ASIM. The FST is the deciding body on:

– Call for, and selection of, research projects submitted by international groups.
– Access and use of data made available from the ASIM Science Data Centre (ASDC), subject to ESA rules on data policy (Human Spaceflight Data Policy 2005).
– Changes of instrument modes and instrument parameters.

The operational- and data access policies are decided by the FST and approved by ESA. There are four members of the FST. They are representatives of the main instrument teams, and of the TARANIS team. The scientific groups that have access to ASIM data form the ASIM Topical Team (ASIM-TT) and may receive travel support if they are within ESA member state countries.

The ASIM Science Data Center (ASDC) receives raw data from ESA's Belgian User Support and Operations Centre (B.USOC) that includes all ASIM instrument and payload data, and selected parameters of the ISS. The data are calibrated and classified according to event type, and the performance of the instruments is monitored. The ASDC also receives measurements from other sources that are used in scientific studies such as lightning data from ground-based networks and cloud properties detected from spacecraft. The data are made available to the ASIM Topical Team and to the instrument teams for technical support of instrument functionality in space. In addition, the ASDC serves as a focal point of the ASIM project, supporting science collaborations, workshops and conferences.

The ASDC is distributed between the University of Bergen, Norway, the University of Valencia, Spain, and DTU Space, Denmark, which is the main node. The primary responsibility of the University of Bergen is to support the operations and calibration of the MXGS high- and low-energy detectors. The University of Valencia supports instrument operations and estimate imaging parameters for angle-of-arrival of TGF photons. DTU Space is responsible for MMIA commissioning, operations and calibration.

The ASDC supports and monitors instrument parameter values, proposes parameter changes and communicates these to B.USOC, subject to approval by the FST. Likewise, the FST may decide on parameter changes based on inputs from the ASIM Topical Team or from the instrument teams, including changes of instrument modes or instrument parameters. The organisation of the ASDC and its main functions are shown in Fig. 4.

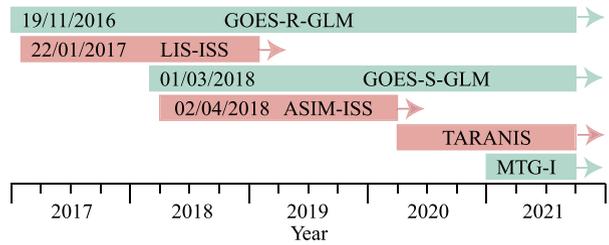

**Fig. 5** Satellite missions that measure thunderstorm electrical activity: blue: geostationary monitoring satellites in geostationary orbit; red: low-earth orbit research missions

To participate in ASIM, project proposals should be submitted to the FST. The ASDC will implement access to the data if a project is approved and the proposers sign an agreement on the use of data. In general, research projects are expected to require access to a subset of ASIM data as well as benefiting the overall ASIM scientific endeavor. For instance, by contributing data from relevant ground-based instruments, scientific expertise, or modeling and laboratory experiments. Groups from all countries are welcome to submit projects; participation is not dependent on a country's membership of ESA. More detailed information on how to participate is given on the ASDC main website http://asdc.space.dtu.dk/.

## 5 Perspectives

The next decade will see the launch of an unprecedented suite of instruments for measurements of thunderstorms, lightning and related atmospheric processes of the atmosphere. For the first time, European and American geostationary meteorological satellites will provide continuous observations of lightning in their respective geographic sectors while three research missions in low Earth orbit will measure new aspects of discharges with dedicated instrumentation. The meteorological monitoring missions are the American GOES-R satellite with the Geostationary Lightning Mapper (GLM) and the European MTG satellite with the Lightning Imager (LI). The GLM and LI instruments are optical cameras that measure lightning day and night. They build on the heritage of the successful Lightning Imaging Sensor (LIS) on the TRMM satellite and the Optical Transient Detector (OTD) on the MicroLab-1 satellite.

The research missions are the ISS-LIS, which is a flight spare of LIS that is installed on the ISS, ASIM—also on the ISS, and the French satellite TARANIS. The ASIM and TARANIS missions carry, for the first time, a suite of dedicated instruments for observations of thunderstorms and their manifestations in the upper atmosphere, characterizing the particle, photon and electromagnetic radiation in wide bands. The time line of the missions is shown in Fig. 5.

ASIM and TARANIS have cameras and photometers that measure the optical properties of lightning and of high-altitude lightning. They include the main lightning band 777.4 nm, which is the band used by the lightning cameras LI (MTG), GLM (GOES-R, S) and ISS-LIS. ASIM and TARANIS also measure x- and gamma-rays and TARANIS additionally detects electromagnetic waves and energetic electrons. The monitoring missions measure clouds, humidity and atmospheric constituents with unprecedented accuracy. GOES-R also carries instruments to measure solar radiation and in-situ plasma and field properties relevant for space weather observations. More information on the instruments and missions are given in Østgaard et al. (2019), Chanrion et al. (2019), Lorenzini et al. (2017), Lefeuvre et al. (2008), Sarria et al. (2017), Cecil et al. (2014), Goodman Stephen et al. (2013), Mach et al. (2007).

It appears that the next 10 years will offer exciting opportunities for advances in the field of thunderstorm research. To harvest the potential of the data, it will be necessary to forge collaborations between the monitoring and science communities. For example, different sensors measure different manifestations of a process, and a combination of measurements leads to new insights into the process. We also note that inter-comparison of data from the same type of sensor, as the 777.4 nm camera common for all missions band—but with different temporal and spatial resolution, will lead to a better understanding of the data. Finally we suggest that a research-based understanding of the physics behind the data will improve their applications.

**Acknowledgements** ASIM is a mission of ESA's SciSpace Programme for scientific utilization of the ISS and non-ISS space exploration platforms and space environment analogues. We would like to thank our many friends and colleagues that have supported the development of the mission. At DTU Space we want in particular to thank Drs. Stephen B. Mende, Harald Frey and Hugh Christian, who have given us invaluable technical advice and unwavering moral support, and our close collaborators of the TARANIS team. In addition we thank the Danish Ministry of Higher Education and Science who supported ASIM from the Danish Globalization Fund for Climate Initiatives (2009–2012) via a special contribution to ESA. We give special thanks to Cecilie Tornøe of the ministry, who skillfully nursed the ASIM payload through the political process, and to ESAs HME Programme Board for supporting the mission. Development of the ASIM instrument computer was supported by the ESA PRODEX contracts PEA 4000105639 and 4000111397. The ASDC is supported by PRODEX contract PEA 4000115884. Additional Danish support was provided by DTU Space and Terma. At the University of Bergen we thank for the Norwegian contribution through ESA programs. Funding at the University of Bergen, for the HED layer was supported by PRODEX contract 4000102100. Additional funding came from Norwegian Research Council (184790/V30, 197638/V30, 223252) and the University of Bergen. Funding at the University of Valencia and INTA, for the MXGS development, integration and test, was supported by the Plan Nacional del Espacio grants: AYA2011_29936_C05, ESP2013_48032_C05 and ESP2015_69909_C05. The ASDC participation is supported by the grant ESP2017_86263_C04.

**Publisher's Note** Springer Nature remains neutral with regard to jurisdictional claims in published maps and institutional affiliations.